\documentclass[
    ,final            
  ]
  {aipproc}

\layoutstyle{6x9}

\begin{document}

\title{Parameter Estimation from an Optimal Projection in a Local Environment}


\author{A.~Bijaoui}{
  address={Observatoire de la C\^ote d'Azur BP4229\\ 06304 Nice Cedex 4, France}
}

\author{A.~Recio-Blanco}{
  address={Observatoire de la C\^ote d'Azur BP4229\\ 06304 Nice Cedex 4, France}
}

\author{P.~de~Laverny}{
  address={Observatoire de la C\^ote d'Azur BP4229\\ 06304 Nice Cedex 4, France}
}

\begin{abstract}
The parameter fit from a model grid is limited by our capability to reduce the number of models, taking into account the number of parameters and the non linear variation of the models with the parameters. The Local MultiLinear Regression (LMLR) algorithms allow one to fit linearly the data in a local environment. The MATISSE algorithm, developed in the context of the estimation of stellar parameters from the Gaia RVS spectra, is connected to this class of estimators. A two-steps procedure was introduced. A raw parameter estimation is first done in order to localize the parameter environment. The parameters are then estimated by projection on specific vectors computed for an optimal estimation. The MATISSE method is compared to the estimation using the objective analysis. In this framework, the kernel choice plays an important role. The environment needed for the parameter estimation can result from it. The determination of a first parameter set can be also avoided for this analysis. These procedures based on a local projection can be fruitfully applied to non linear parameter estimation if the number of data sets to be fitted is greater than the number of models.
\end{abstract}

\maketitle

\section{Models and Parameters.}

The comparison between observations and models obtained from simulations is nowadays a common activity in astrophysics. Each model is defined by a set of physical parameters $\Theta$. It carries out a set of data which can be reduced to equivalent observational measurements $S$, taking into account the properties of the used instrumentation. Detector noise and instrumental uncontrolled variations lead to associate to observational data $O$ a conditional probability density function (PDF) $p(O/S)$.  Currently, this PDF is admitted to be Gaussian.

The application of the Maximum Likelihood (ML) principle \cite{KS} leads to get the parameter values $\hat{\Theta}$ which maximize $p(O/S)$. Taking into account the hypothesis of a Gaussian noise, this is equivalent to determine the parameters which minimize a quadratic distance between the observations and their simulations. A large literature exists on the least mean squares (LMS) method \cite{LMS}. In case of an analytical model, normal equations are derived and the problem is reduced to a classical numerical problem. Generally, the models carry out equivalent observational data which cannot be expressed as analytical functions of the physical parameters. Each comparison needs to compute a model, which can be very CPU time consuming. In order to reduce the cost, the models are computed on a parameter grid, its sampling being as large as possible such that the parameter extraction results from an interpolation in this space.

In the framework of the determination of the atmospheric parameters from the stellar spectra obtained with the Gaia RVS \cite{RVS} we developed a new algorithm called MATISSE \cite{ABdL}. This algorithm is based on the determination of the parameter set by projection of the observed spectrum $O$ to a set of vectors $B_i$ computed for each parameter $\theta_i$. MATISSE belongs to a class of statistical methods which allows one to estimate parameters from of an optimal projection on a local environment. The projection vectors derive from a multilinear regression \cite{LLR} which takes only into account the models locally computed.

\section{The distance minimization and the MATISSE method.}

The parameter estimation problem consists into finding the minimum distance $D(\Theta)$ between the observable $O$, here an observed spectrum, and a peculiar model from a theoretical grid, here a theoretical spectrum $S$. In order to simplify the presentation it is admitted that the noise variance is constant and that the measurements are independent. Therefore, the considered distance is:
\begin{equation}\label{eq:02.3}
  D(\Theta)=\sum_{l=1,L} [O(l)-S(l,\Theta)]^2.
\end{equation}

The resulting normal equation for a given parameter $\theta_i$ can be derived. The related equation system is generally non linear. To solve it, one needs to start from a given parameter set $\Theta^{(0)}$ and to linearize it around these values. The corrections can be large so that an iteration could be necessary. There is no guarantee that the values after convergence correspond to the absolute minimum distance. The raw solving way would consist into computing a set of models with a sufficient grid resolution for avoiding non convexity of the distance function in a given mesh. All the distances to the models are computed and the absolute minimum is localized. In a convex region the distance function is quite a quadratic function of the parameters and a simple linearization is sufficient. However, for models associated to many parameters which can vary on a large dynamical range, this procedure could be too heavy. Many algorithms were proposed for avoiding the computation of the whole model grid. Since the 70's, genetic algorithms were recognized as a fine way for solving optimization problems \cite{GA}. This technique was applied for stellar parameter estimation from spectrograms \cite{AP}. Simulated annealing \cite{SA} was also developed and appeared to be well adapted to carry out parameter estimation \cite{PESA}. Artificial neural networks (ANN) were also used for this specific problem \cite{BJ}.

Let us come back to the normal equations. The parameter corrections can be written as:
\begin{equation}\label{eq:02.6}
   \sum_{l=1,L} {\partial S(l,\Theta^{(n)})\over \partial
  \theta_i}[O(l)-S(l,\Theta^{(n)})]=
  \sum_{i'=1,I} \delta\theta_{i'} \sum_{l=1,L}
  {\partial S(l,\Theta^{(n)})\over \partial
  \theta_i}{\partial S(l,\Theta^{(n)})\over \partial
  \theta_{i'}}.
\end{equation}
So we can write the relation:
\begin{equation}\label{eq:02.7}
  \delta\Theta=G^{-1} [T (O-S)].
\end{equation}
where $O-S$ designs the residual vector (length $L$), $T$ the $L\times I$ model derivative matrix and $G$ the $I\times I$ Gram-Schmidt matrix of the model derivatives. With Equation (\ref{eq:02.7}), the parameter corrections are expressed as linear combinations of the model residuals. The associated coefficients depend on the partial derivatives. In our studied problem only computed models on a grid are available. Approximated derivatives can be computed but it is better to embed this approximation into a more general scheme taking into account the grid sampling. In this scheme, the MATrix Inversion for Spectral SynthEsis (MATISSE \cite{ABdL}) method, the algorithm determines a vector, $B_i$, allowing one to derive a particular parameter $\theta_i$ by projection of an input observed spectrum on it. The $B_i$ vector is derived from an optimal linear combination of models in a local environment, the environment depending on the model linearity versus the parameters. $B_i$ is determined from the principle that the statistical correlation between the input and the output parameter values is maximum. With a spectra energy normalization this is equivalent to minimize the distance between the input and the output parameter values.

First of all, the data on a particular $\theta_i$ parameter and the spectra of the grid are subtracted by their mean values. The $B_i$ vector is then constructed from:
\begin{equation}\label{eq:3.1}
B_i(l) = \sum_{j=1,J} ~\alpha_{ij}  S_j(l).
\end{equation}
$\theta_i$ is estimated by the spectrum projection:
\begin{equation}\label{eq:3.2}
\hat {\theta_i}  = \sum_{l } ~B_i(l)O(l).
\end{equation}
In other words, $\Theta$ is estimated by multilinear regressions done on the spectrum values. Equation (\ref{eq:3.1}) is applied to all the model spectra $S_j(l)$. Combining with Equation (\ref{eq:3.2}) we obtain :
\begin{equation}\label{eq:3.31}
{\Theta}_i  = C \alpha_i,
\end{equation}
where $C=[c_{jj'}]$ is the correlation matrix of the spectra and $\Theta_i$ the vector of the parameter $i$ for all the spectra. For an invertible variance-covariance matrix, we get:
\begin{equation}\label{eq:3.71}
\alpha_i = C^{-1} \Theta_i,
\end{equation}
which leads for the training set to:
\begin{equation}\label{eq:3.72}
  {\hat{\Theta}}_i= SB=S(S\alpha_i)=CC^{-1}\Theta_i=\Theta_i.
\end{equation}

The exact values are restored. In order to be sure that a peculiar model was not forgotten for the basis building a redundancy is necessary. That leads to a non invertible matrix. The LMS solution between the input parameters and their estimated ones leads to:
\begin{equation}\label{eq:3.61}
C^T C\alpha_i =  C^T\Theta_i.
\end{equation}
Even if $C$ is singular, the multiplication by its joint matrix $C^T$ brings a solution in the LMS sense. The inversion can be done, for example, using the classical Landweber iterative algorithm (Landweber, 1951).

\begin{figure*}[htb]
   \centering
   \includegraphics[width=12.0cm]{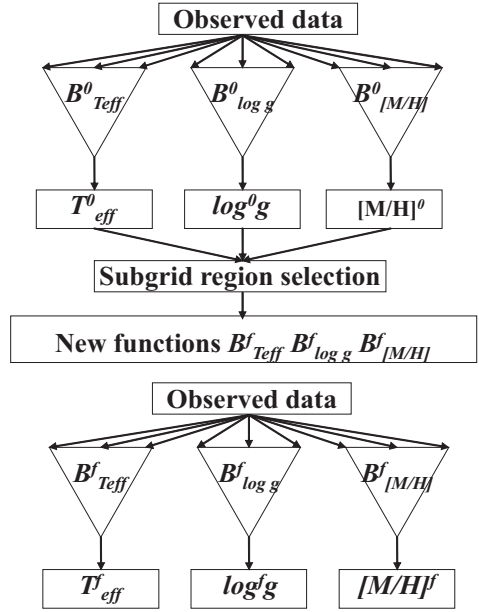}
   \caption{Scheme of the MATISSE algorithm for the estimation of the
   stellar parameters.}
              \label{Scheme}%
    \end{figure*}

The multilinear regression model stands only if the model is quite linear. Generally it is not the case and it is needed to process in two steps (Fig. \ref{Scheme}):
\begin{enumerate}
\item A preliminary estimation of the parameters is performed through the use of $B_\Theta^{(0)}$ functions computed from spectra combinations spanning the parameter range of the whole grid.
\item This first estimation allows one to focus into a subregion of the model grid, where new $B_{\Theta}^{(f)}$ functions constructed from a subset of theoretical spectra are applied.
\end{enumerate}

\section{The Objective Analysis with kernel.}

The {\em Objective Analysis} recovers a large set of interpolation methods developed in the framework of Geosciences. The function $f(x)$ is known on J points ($x_j$, $j\in (1,J)$) and it is interpolated with the rule:
\begin{equation}\label{13.5}
    f(x)={\sum_j w_j(x) f(x_j)\over \sum_j w_j(x)}
\end{equation}

Many determinist or statistical methods, like the kriging  \cite{GM} \cite{Kr}, were proposed for the determination of the coefficients $w_j(x)$. The weights can be directly obtained from a Kernel function $K({x\over a})$ \cite{OA}, $a$ being a scaling parameter. The Nadaraya-Watson weighted average \cite{NW} consists in the following rule:
\begin{equation}\label{13.7}
    \hat{f}(x)={\sum_j K({|x-x_j|\over a}) f(x_j)\over \sum_j K({|x-x_j|\over a})}.
\end{equation}
In order to obtain a true interpolation for which the values are exactly recovered at the mesh points, the set $\overline{f}(x_j)$ such that:
\begin{equation}\label{13.8}
    f(x_j)={\sum_{j'} K({|x_j-x_{j'}|\over a}) \overline{f}(x_{j'})\over \sum_{j'} K({|x_j-x_{j'}|\over a})}
\end{equation}
is determined. Equation (\ref{13.8}) is solved iteratively. The interpolation rule on the parameters $\theta_i$ is thus:
\begin{equation}\label{13.9}
    \hat{\theta}_i(O)={\sum_j K({|O-S_j|\over a}) \overline{\theta}_{ij}\over \sum_j K({|O-S_j|\over a})},
\end{equation}
where the $\overline{\theta}_{ij}$ are previously determined by inversion.

Let us consider the Epanechnikov kernel \cite{Epa}:
\begin{equation}\label{4.1}
    K(x)={3\over 4}(1-x^2) \qquad {\rm if}\qquad |x|\le 1 \qquad {\rm otherwise} \qquad K(x)=0.
\end{equation}
 Taking into that the spectra normalization the weights can be simplified as:
\begin{equation}\label{4.3}
    w_{jj'}=H(c_{jj'}-c),
\end{equation}
where $c$ is a correlation threshold and $H(x)$ the Heaviside function ($H(x)=1$ if $x\ge 0$ and null otherwise). The  weights for a spectrum $O$ are thus:
\begin{equation}\label{4.4}
    W_{j}=H(c_{j}-c)=H(\sum_l O(l)S_j(l)-c).
\end{equation}

The parameter $\theta_i$ is estimated by the relation:
\begin{equation}\label{4.6}
    \hat{\theta}_i={\sum_j H(\sum_l O(l)S_j(l)-c) \overline{\theta}_{ij}\over \sum_j H(\sum_l O(l)S_j(l)-c)}.
\end{equation}
Defining $e(O)$ as the set of spectra $S_j$ such that $c_{j}> c$, relation (\ref{4.6}) can be written as:
\begin{equation}\label{4.7}
    \hat{\theta}_i={\sum_l O(l)[\sum_{j\in e(O)} S_j(l)\overline{\theta}_{ij}]-c\sum_{j\in e(O)} \overline{\theta}_{ij}\over W},
\end{equation}
where $W=\sum_{j\in e(O)} W_j$. Thus it results that:
\begin{equation}\label{4.8}
    \hat{\theta}_i=\sum_l O(l)B_i^e(l)-\theta_i^e
\end{equation}
where
    $B_i^e(l)={1\over W}\sum_{j\in e(O)} S_j(l)\overline{\theta}_{ij}$
and
    $\hat{\theta}_i^e={c \sum_{j\in e(x)} \overline{\theta}_{ij}\over W}$.
The parameters $\overline{\theta}_{ij}$ are determined by the inversion of the equation:
\begin{equation}\label{4.11}
    \theta_{ij}={\sum_{j'} H(c_{jj'}-c) \overline{\theta}_{ij}\over \sum_j H(c_{jj'}-c)}.
\end{equation}

This equation, with a null threshold, is equivalent to the one associated to the MATISSE method, giving the $\alpha_i$ value (Equation \ref{eq:3.31}).  In the MATISSE case, the environment is defined from the grid sampling. Here the environment results from the correlation sampling for the objective analysis. The estimation is done by a correlation product, with a function which varies according to the correlations. It can be noted that the $B_i^e$ varies for a spectrum to another, but the contributions associated to each spectrum are computed in one step.

\section{Experimentations on a spectra set.}

A set of $1386$ spectra with $971$ Gaia RVS spectral elements were computed. In this grid, $3$ physical parameters vary, the temperature $T$, the surface gravity $\log g$ and the metallicity $\mu$. Experiments were done with different kernels. In the case of the Epanechnikov kernel, different correlation thresholds were examined. For the studied grid, the environment, defined by the number of remaining spectra for each one, seems to be sufficient at a $0.995$ threshold. In Table \ref{tab:a} a comparison is presented with different experiments:
\begin{description}
  \item[Kernel] The parameters are restored using the Nadaraya-Watson weighting average, with the Epanechnikov kernel. The correlation threshold is equal to $0.995$. The environment varies from $2$ to $68$ spectra with this value. The minimum number is obtained at points on the grid bound, for which the environment is insufficient to get an accurate value.
  \item[MATISSE B0] The parameters are determined by one projection. The maximal errors are greater than the grid steps. This operation allows one only to localize the environment for obtaining the correct values.
  \item[MATISSE B0+Bf] After localization, a projection is done for obtaining correct values. Few iterations may be needed in order to get the environment corresponding to the parameters.
  \item[MATISSE with rejection] The previous algorithm works perfectly on $99$\% of the grid spectra. The errors are related to the grid boundary. At the first step the estimated parameters may be too far from real ones. The $B_f$ corrections do not allow us to converge correctly. A simple test based on the correlation between the restored spectrum and the observed one allows us to reject the bad estimations.
\end{description}
\begin{table}
\begin{tabular}{lrrrrrr}
\hline
  \tablehead{1}{c}{b}{Method}
  & \tablehead{1}{c}{b}{$T$-max}
  & \tablehead{1}{c}{b}{$\sigma(T)$}
  & \tablehead{1}{c}{b}{$\log g$-max}
  & \tablehead{1}{c}{b}{$\sigma(\log g)$}
  & \tablehead{1}{c}{b}{$\mu$-max}
  & \tablehead{1}{c}{b}{$\sigma(\mu)$}
    \\
\hline
{Kernel at $0.995$}  &  122 &  13 & 0.42  & 0.06 & 0.17 & 0.03\\
{MATISSE  B0}    & 1402 & 451 & 2.95  & 0.69 & 1.32 & 0.26\\
{MATISSE B0+Bf} & 1709 & 104 & 6.    & 0.19 & 1.34 & 0.08\\
{B0+Bf 99\% stars}&    0. &   0. & 0.     & 0.   & 0.   & 0.  \\
\hline
\end{tabular}
\caption{Experimental results on a reduced grid.}
\label{tab:a}
\end{table}

These experiments show that the kernel method is well designed for a first estimation of the parameter values on a reduced grid. It is necessary for that to compute theoretically all the correlations. Specific algorithms have to be developed for reducing this number taking into account the correlation threshold.

%

\section{Conclusion.}
The MATISSE method allows one to improve model fitting in the case of:
\begin{itemize}
  \item A capability to describe observations by a suitable model depending of few parameters;
  \item The models can not be analytically manipulated, avoiding to use their derivatives for the estimations;
  \item The need to fit a huge number of observational data.
\end{itemize}

The main part of the computation is related to the basis estimation, the parameter estimation itself being very fast. The use of the Epanechnikov's kernel method allows one to get in one inversion all the coefficients. The number of spectra having a correlation above the threshold according to a given grid point indicates if locally the sampling is in agreement with the searched accuracy.

Taking into account their capabilities to process many sets of astronomical observations, with a high statistical consistence level, the estimation methods based on local projections  like MATISSE open a new window on astrophysical modelization for large astrophysical surveys.

\end{document}